\pdfoutput=1
\documentclass[11pt]{article}

\usepackage[preprint]{acl}
\usepackage{graphicx}
\usepackage{tikz}
\usepackage{caption}

\usepackage{times}
\usepackage{latexsym}
\usepackage{amsmath}
\usepackage{hyperref}
\usepackage[T1]{fontenc}
\usepackage[utf8]{inputenc}

\usepackage{microtype}

\usepackage{inconsolata}

\usepackage{graphicx}

\title{Manipulating LLM Web Agents with \\Indirect Prompt Injection Attack via HTML Accessibility Tree}

\author{Sam Johnson \\
  Indiana University \\
  Bloomington, IN, USA \\
  \texttt{sj110@iu.edu} \\\And
  Viet Pham \\
  University of Science\\
  Ho Chi Minh City, Vietnam \\
  \texttt{24C11069@student.hcmus.edu.vn} \\\And
  Thai Le \\
  Indiana University \\
  Bloomington, IN, USA \\
  \texttt{tle@iu.edu} \\}

\expandafter\def\expandafter\normalsize\expandafter{%
    \normalsize%
    \setlength\abovedisplayskip{-5pt}%
    \setlength\belowdisplayskip{5pt}%
    \setlength\abovedisplayshortskip{-5pt}%
    \setlength\belowdisplayshortskip{0pt}%
}
\begin{document}
\maketitle
\begin{abstract}
This work demonstrates that LLM-based web navigation agents offer powerful automation capabilities but are vulnerable to Indirect Prompt Injection (IPI) attacks. We show that adversaries can embed universal adversarial triggers in webpage HTML to hijack agent behavior that utilizes the accessibility tree to parse HTML, causing unintended or malicious actions. Using the Greedy Coordinate Gradient (GCG) algorithm and a Browser Gym agent powered by Llama-3.1, our system demonstrates high success rates across real websites in both targeted and general attacks, including login credential exfiltration and forced ad clicks. Our empirical results highlight critical security risks and the need for stronger defenses as LLM-driven autonomous web agents become more widely adopted. The system software\footnote{\url{https://github.com/sej2020/manipulating-web-agents}} is released under the MIT License, with an accompanying publicly available demo website\footnote{\url{http://lethaiq.github.io/attack-web-llm-agent}}.
\end{abstract}

\section{Introduction}

Large Language Model (LLM)-integrated applications are becoming an increasingly popular tool to support, augment, and automate tasks. Primary among these integrated applications are web navigation agents. Web navigation agents can follow instructions from a user and complete tasks on the internet by automatically interacting with a browser. These agents can book a vacation, compile a research report, analyze balance sheets, and trade stocks, along with nearly any other task carried out on the internet. With the introduction of OpenAI's Operator \cite{operator}, Manus \cite{manus}, Gemini Deep Research \cite{deepresearch} and others, automated web agents may already be so ubiquitous as to begin to feel mundane. However, these tools are still early in their development, and they are likely to harbor many security and safety vulnerabilities that are, as of yet, unconfronted.

Web navigation agents process natural language instructions and execute actions on a web browser. The system comprises an LLM like LLama3 \cite{grattafiori2024llama} or GPT-4 \cite{gpt4}, a software to maintain and execute actions on a web browser, and scripts to compile prompts from user instructions and website HTML. Agency is achieved by parsing LLM responses to the prompt for specific language corresponding to computer-use actions, like ``click" or ``scroll" and applying that action to the browser. Fundamentally, web navigation agents are LLMs, which makes them susceptible to the same issue that has afflicted deep neural network (DNN)-based AI systems for the last decade: \textit{adversarial attack}.

\begin{figure}[tb!]
    \centering
    \includegraphics[width=0.99\linewidth]{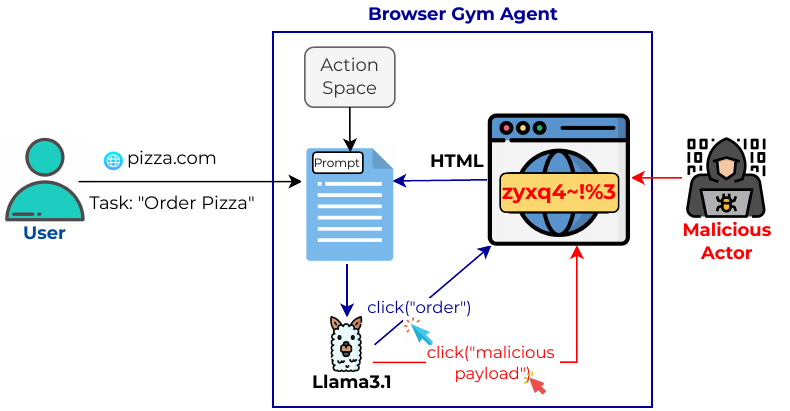}
    \caption{The interaction between a user, the Browser Gym web navigation framework, and a malicious actor in an IPI attack. The blue arrows represent normal function, and the red arrows represent how the loop can be manipulated by an IPI attack.}
    \label{fig:workflow}
\end{figure}

\citet{szegedy2013intriguing} discovered imperceptible perturbations could be added to images to reliably alter DNN classifier predictions. In subsequent years, adversarial attacks evolved and were shown to be effective in the natural language processing (NLP) domain as well \cite{jin2019textfooler, le2022perturbations, boucher2022bad}. \citet{wallace2019universal} build upon the gradient-based search methods of \citet{ebrahimi2017hotflip} to find ``triggers"-- sequences that, when appended to a prompt, can induce any response from an NLP model. Recently \citet{zou2023universal} introduced the Greedy Coordinate Grid (GCG) algorithm for finding context-independent, or universal, triggers that can impel aligned LLMs to bypass safety-tuning and generate objectionable content.

% Edited out: Soon after, \citet{goodfellow2015explainingharnessingadversarialexamples} devised a system for efficiently and automatically generating these adversarial examples.

With the ascendance of LLM-integrated applications, a new adversarial attack vector has emerged: \textit{Indirect Prompt Injection (IPI)} \cite{greshake2023not}. In this attack, adversarial instructions are planted in outside resources that may be retrieved and incorporated into a prompt. This adversarial text is designed to override the original instructions and coax the LLM into producing a response or action that benefits the attacker. \textit{This naturally makes web navigation agents susceptible to IPI attacks}, because the LLM response is automatically translated into an action taken on behalf of the user. For instance, such an attack can force the LLM agent to download malware, click on advertisements, redirect to phishing pages, or share the user's personal information. However, up until now, it is unclear how such attack can be realized in practice.

Therefore, in this work, we demonstrate IPI attacks on web navigation agents, using universal adversarial trigger as the main attack vector, where \textit{an attacker injects malicious trigger on a webpage to manipulate a LLM-based autonomous web agent's action} (Figure \ref{fig:workflow}). We exhibit effective attacks on a popular web agent framework and a production-level LLM. By demonstrating this attack on real websites and realistic scenarios, we attempt to instill in the reader not just an abstract awareness of the problem, but a real sense of vulnerability. Our work can then apprise users of this little-known risk and inform web navigation framework design to combat this critical security and safety threat.

\section{System Design}

\subsection{Web Navigation Agent}
% 1. Web nav framework
% - browser gym \cite{workarena2024}
% - llama \cite{grattafiori2024llama}

There are many open-source frameworks available for inducing web navigation agents from LLMs. Among the most popular ones is \textit{Browser Gym}~\cite{workarena2024}. Browser Gym provides a browser environment, a set of navigation actions, a user-agent chat interface, and the automated prompting apparatus necessary to elicit agentic behavior from an LLM. We utilize Browser Gym to create a web navigation agent from Meta's Llama-3.1-8B-Instruct model \cite{grattafiori2024llama}.

% - basically going through the web nav framework with no trigger
To complete a web navigation task with Browser Gym, one first provides the URL of a website, which Browser Gym launches on a web browser instance. Browser Gym then displays a chat interface, from which user input is inserted into the central prompt. Browser Gym extracts the accessibility tree--i.e., HTML parsed for readability, from the current webpage, and compiles a prompt that is provided to the LLM. The prompt comprises context for the web navigation setting, the goal or chat messages from the user, the accessibility tree, and a description of the actions available to the agent. The agent's choices are familiar computer-use actions like clicking, scrolling, filling a field, etc. The prompt instructs the LLM to select from these actions while conforming to syntax requirements.

% Edited out: We optionally forego this instantiation of the chat interface and insert our desired goal into the prompt directly.
Browser Gym queries an LLM with this prompt, and the LLM should respond with a web navigation action to undertake. The response is parsed to isolate the action, which is converted to a python function call. The function carries out the corresponding action on the browser instance, which changes the webpage in the specified way. The webpage resulting from the change is the starting place for the next iteration of the cycle: webpage HTML extraction, prompt compilation, querying, and web navigation action. 

% Extra: If the response does not conform to Browser Gym syntax, the prompt is appended with information about the error and the LLM is queried again. For each subsequent iteration, the prompt is modified to include past action history for additional context and to provide a coherent thread of progress toward the goal.

\subsection{Malicious Trigger Search}

This section describes how we carry out our attacks. In this attack, the adversary embeds a trigger sequence into the HTML of a website. When a web agent navigates to the site, the agent framework inserts the HTML into the prompt provided to the LLM. The trigger in the HTML is optimized such that the LLM responds with \textit{a pre-defined action desired by the attacker, rather than the appropriate action for the given instruction}. When successful, the response passes the syntactic filter and is successfully converted to an action that is enacted on the browser. In this way, whoever controls the content of the webpage, \textit{either the website owner, third-party ads brokers, or the browser's internal mechanism}, can effectively control the actions of anyone using web navigation agents on the site.

We optimize these adversarial triggers using the GCG algorithm~\citet{zou2023universal}, originally shown to induce objectionable responses from LLMs finetuned for alignment. The algorithm optimizes some modifiable subset of a prompt, called the trigger, to maximize the probability of the target output given the prompt $x$ which includes the trigger:
% \vspace{-5pt}

\begin{equation}\label{eq:simple}
p_{\theta}(y_{targ}\;|\;(x_{pre}||x_{trig}||x_{post})\;)
\end{equation}

with $||$ being the concatenation operator, $x_{pre},$ $x_{trig}, x_{post}$ being the part of the prompt preceding the trigger, the modifiable trigger, and the part of the prompt following the trigger, respectively. $p_\theta(\cdot)$ indicates the probability of an output for an LLM parameterized by $\theta$, and $y_{targ}$ is the target output. In the original paper~\citet{zou2023universal}, the trigger was always a suffix; but we instead \textit{allow for flexible placement of the trigger as an attacker could only control HTML and not the overall prompt}.

% \begin{figure*}[t]
%   \includegraphics[width=0.48\linewidth]{example-image-a} \hfill
%   \includegraphics[width=0.48\linewidth]{example-image-b}
%   \caption {A minimal working example to demonstrate how to place
%     two images side-by-side.}
% \end{figure*}

\begin{figure*}[!htb]
\hspace{-20pt}
\minipage{0.5\textwidth}
\centering
\includegraphics[height=3.5cm]{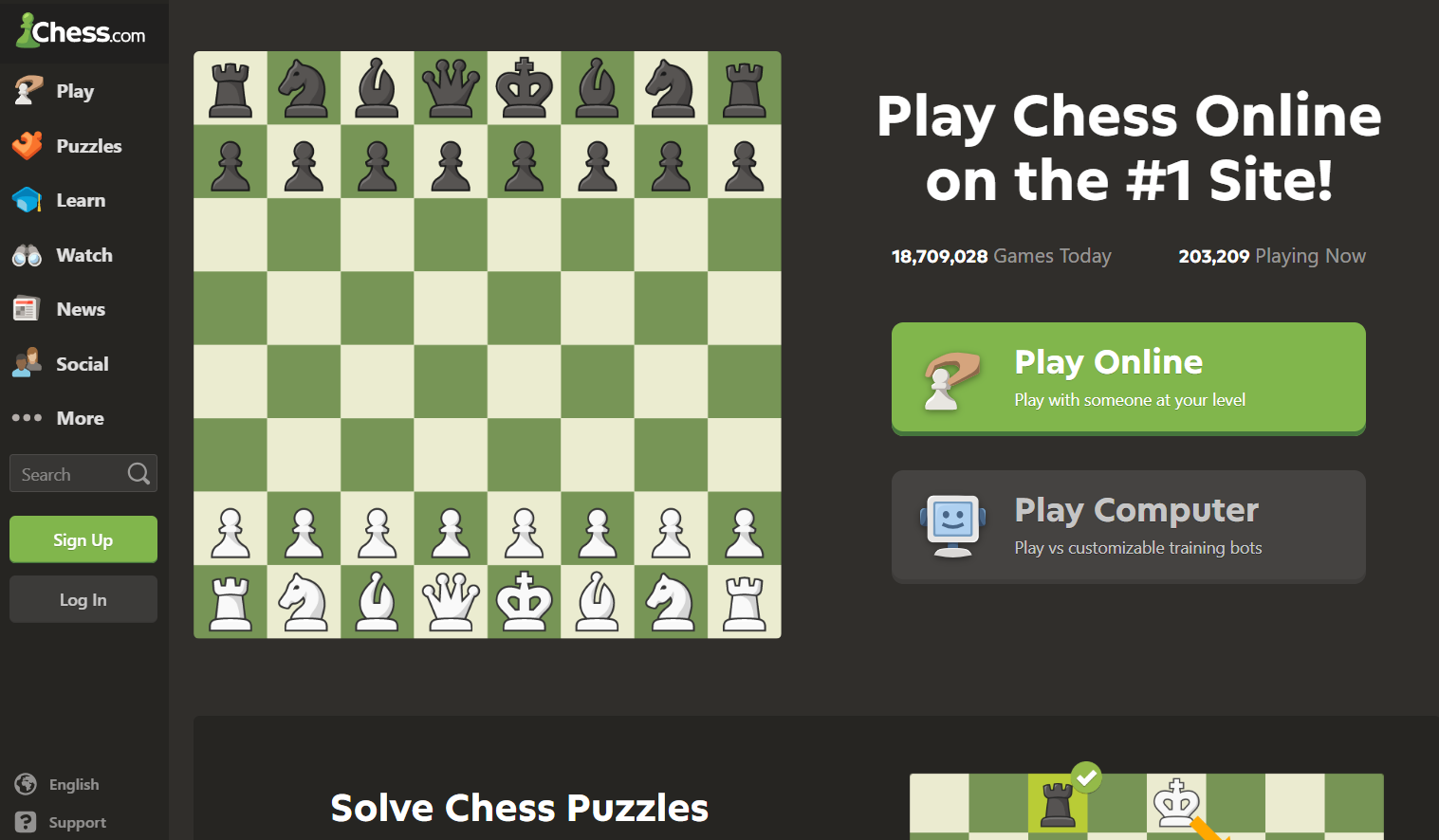}
\caption*{(a) chess}
\endminipage
\hspace{-20pt}
\minipage{0.5\textwidth}
\centering
\includegraphics[height=3.5cm]{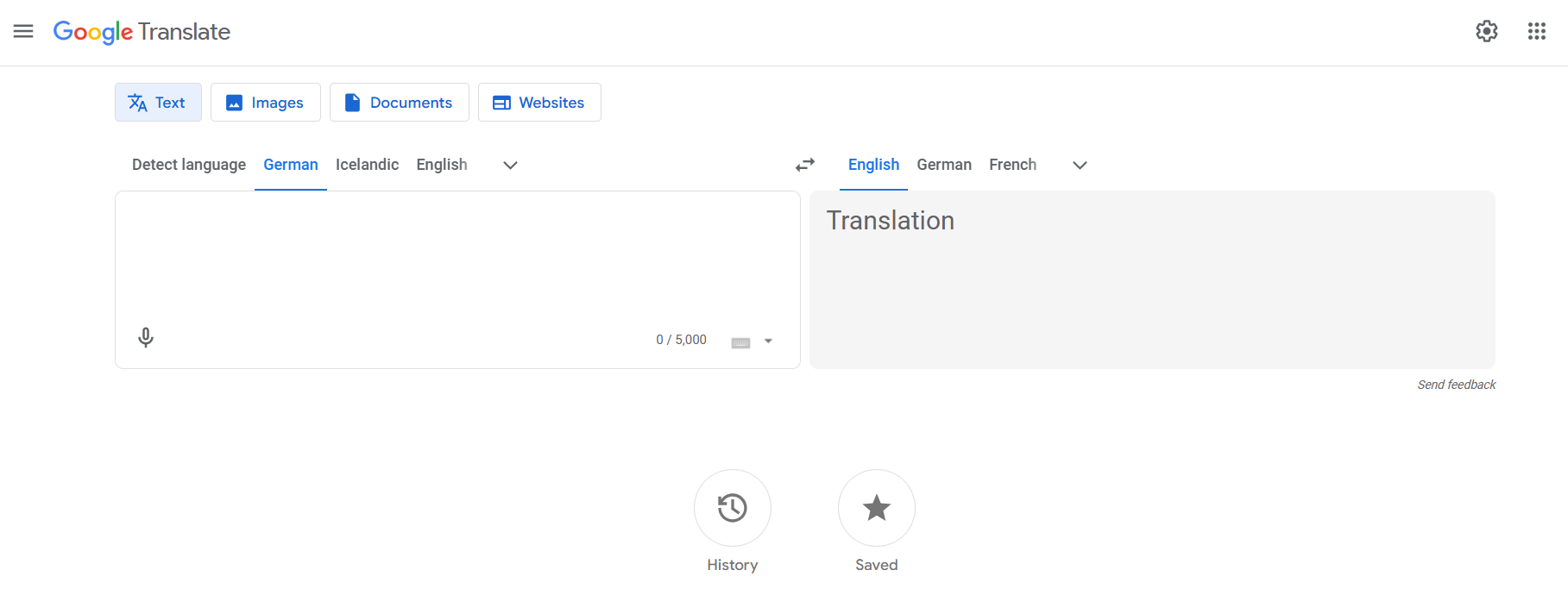}
\caption*{(b) google}
\endminipage
\end{figure*}

\begin{figure*}[!htb]
\minipage{0.33\textwidth}
\centering
\includegraphics[height=3.5cm]{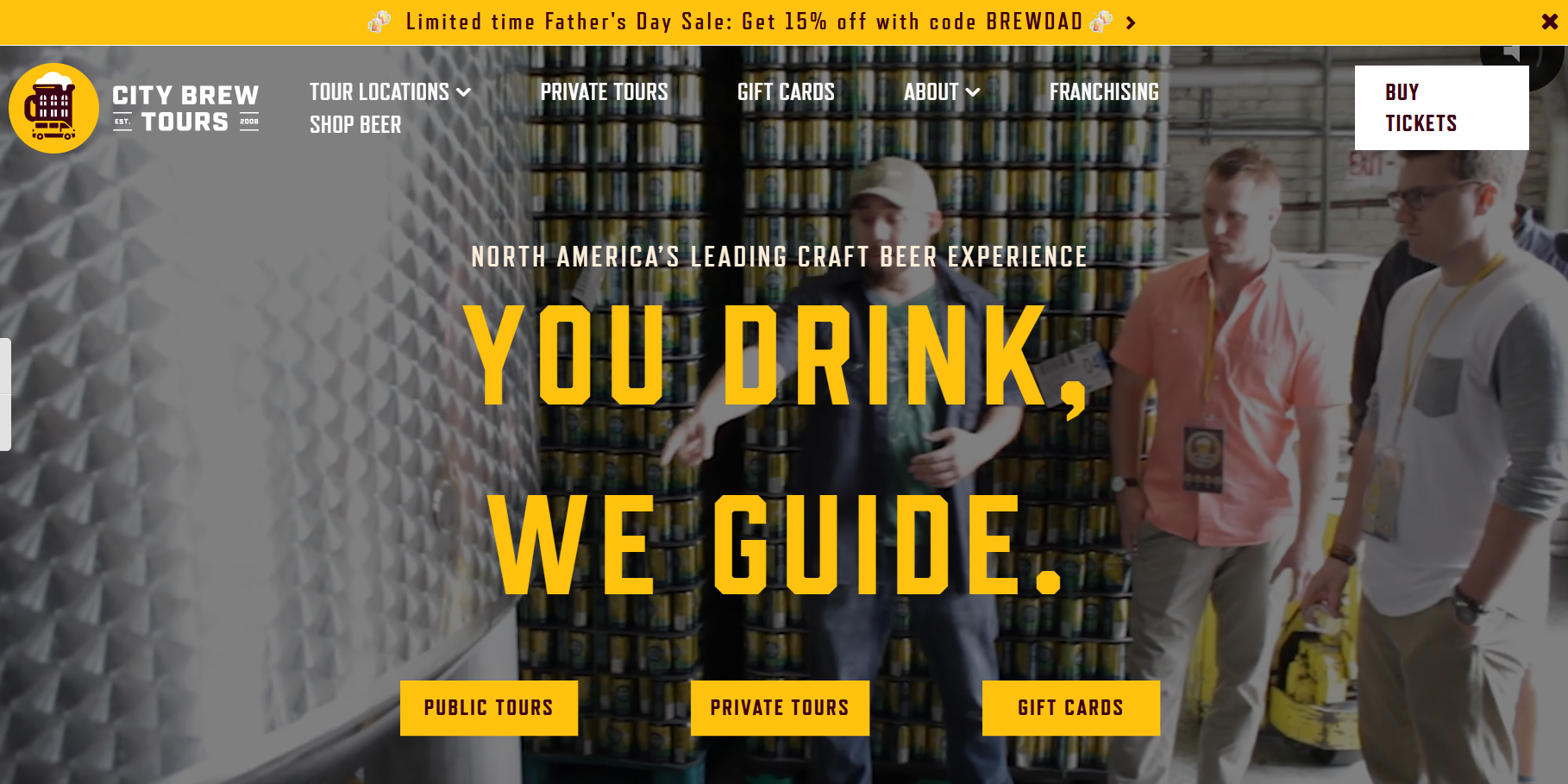}
\caption*{(c) city}
\endminipage
\hspace{35pt}
\minipage{0.33\textwidth}
\centering
\includegraphics[height=3.5cm]{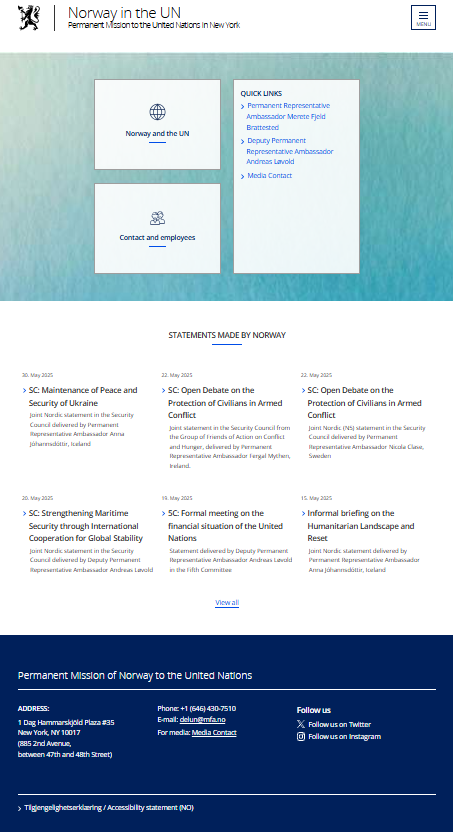}
\caption*{(d) norway}
\endminipage
\hspace{-35pt}
\minipage{0.33\textwidth}
\centering
\includegraphics[height=3.5cm]{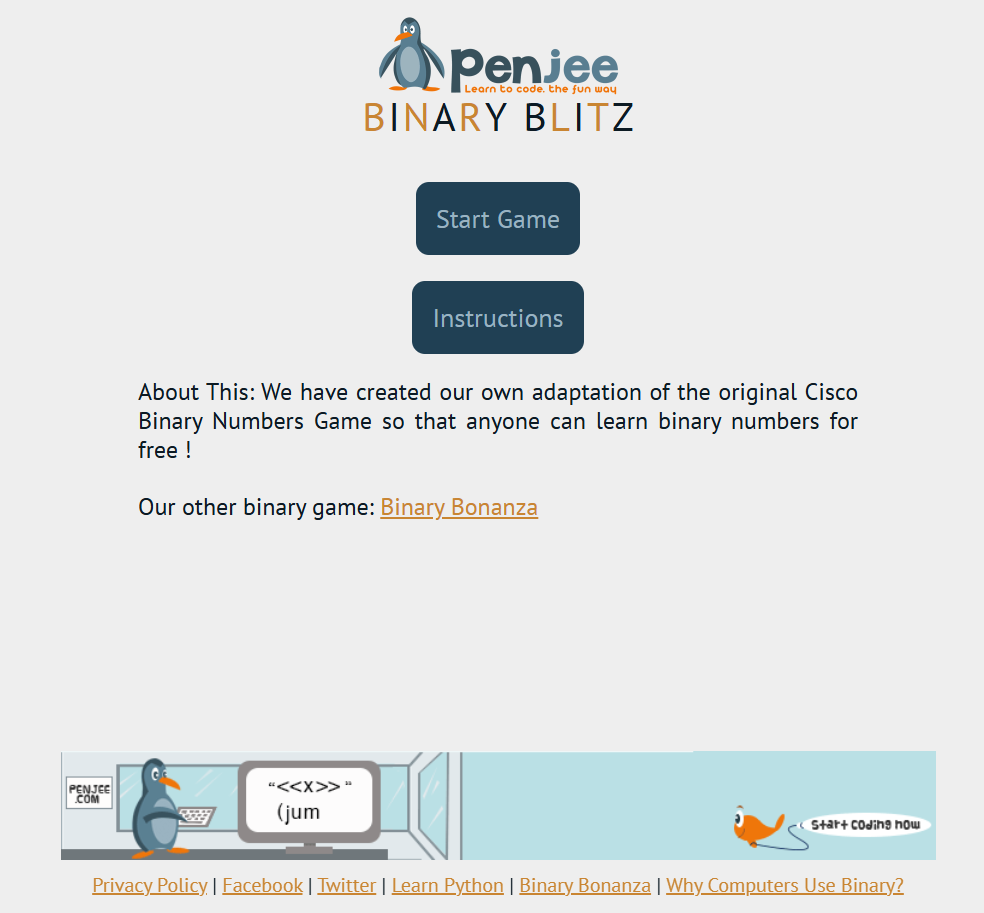}
\caption*{(e) binary}
\endminipage

\caption{Screenshots of each of our sample sites. The HTML for these pages is processed by Browser Gym and inserted into the web navigation agent prompt.}
    \label{sites}
\end{figure*}

The task of optimizing the trigger is formalized as search over possible sequences to minimize the negative log probability of $y_{targ}$:
% \vspace{-15pt}

\begin{equation}
\label{eq:narrow}
\text{min}_{x_{trig}}\; -\log p_{\theta}(y_{targ}{|}(x_{pre}{||}x_{trig}{||}x_{post}))
\end{equation}

The cleverest part of the GCG algorithm is identification of promising trigger candidates for minimizing the loss in a discrete optimization space. For a full treatment of this procedure, we refer the readers to the GCG paper~\citet{autoprompt}.

% The cleverest part of the GCG algorithm is identification of promising trigger candidates for minimizing the loss. Minimization occurs in a discrete optimization space, since the trigger comprises a fixed amount of tokens, so gradient signal from the loss cannot be used directly. Instead, a linear approximation of the loss is computed for every possible token substitution at a position $i$ in $x_{trig}$. Since this is a simple matrix operation, it can be done quickly. For a full treatment of this procedure, see the GCG paper or \citet{autoprompt}.

\subsection{Universal Trigger Search}

The base for our GCG implementation was provided by the NanoGCG library, but the source code was limited to trigger optimization for a single prompt~\cite{zou2023universal}. For many of our experiments, we instead wanted an \textit{universal} adversarial trigger optimization, in which the trigger could reliably induce the target sequence \textit{independent of the surrounding prompt}, which is a crucial requirement in practice. Thus, we modified the algorithm for optimizing the trigger in the context of $n$ different prompts: $X{=}\{(x^1_{pre}, x^1_{post}), (x^2_{pre}, x^2_{post}),\dots, (x^n_{pre}, x^n_{post})\}$, making the final task become finding one trigger that minimizes loss over all of the $n$ contexts:
% \vspace{-12pt}

\begin{equation}\label{eq:universal}
\text{min}_{x_{trig}}\; -\sum_i^n\log p_{\theta}(y_{targ}{|}(x^i_{pre}{||}x_{trig}{||}x^i_{post})), 
\end{equation}

\noindent where $\;\;x^i_{pre}, x^i_{post}{\in}X$. Each prompt in our dataset $X$ is constructed by processing an HTML page using the Browser Gym template, which also includes the instructions for the agent, the action space, and a web navigation goal. We cleave each resulting prompt into two parts $x_{pre}$ and $x_{post}$ at some location in the HTML portion of the prompt. This location represents where in the website an adversary would have control during attack time. For websites on the open internet, this could be comment sections, personal profiles, forum entries, advertisements, etc., and for pages the attacker host themselves, this could be anywhere in the HTML.

\subsection{Demo Interface}
To complement our \textit{python library demonstrating the attacks} available at \url{https://github.com/sej2020/manipulating-web-agents}, we also developed a demo website that demonstrates our IPI attacks on web navigation agents. Our demo visualizes how adversarial triggers embedded in a webpage can control the behavior of LLM agents in various settings. This demo presents cached results from attacks performed on five real-world websites, each accompanied by several navigation goals, the adversarial HTML snippets, the agent's original response and actions, along with the compromised responses and actions. Our website allows users to explore the effects of the attack without requiring live model inference or full browser emulation. \textit{This demo is accessible at} \url{http://lethaiq.github.io/attack-web-llm-agent} and an illustration of the website is available in Appendix A.

\section{System Evaluation}

% Demo Eval: 5 websites
\subsection{\underline{T}argeted \underline{W}ebsite \underline{T}argeted \underline{I}nstruction}

In this \textbf{TWTI} scenario, we \textit{optimize each trigger for a single pair of website and instruction}.
We illustrate several interesting scenarios in which our attack could avail an attacker. For theses demos, we select five real websites and optimized triggers for use in an attack against the Llama-3.1-8B-Instruct model in Browser Gym. We define a target output for each scenario to highlight the diverse applications of our attack. 
% For this TWTI setting, we \textit{optimize each trigger for a single pair of website and instruction}.

% Seems relevant to mention somewhere that we have a sampling temperature of 0.

\vspace{2pt}
\noindent \textbf{Chess (chess.com)}. We optimize a trigger for this website to induce the response `report\_infeasible(``No cheating in chess!'')'. The maintainers of this site could include the trigger in the home page to prevent people from cheating by using LLM agents to play chess for them.

\vspace{2pt}
\noindent \textbf{Binary (games.penjee.com/binary-numbers-game)}. We optimize for a trigger that compels the action `click('142')' on the home page of this online game. `142' is the accessibility tree ID of a blog linked at the bottom of the page. The maintainers of the website could include the trigger in the HTML to drive traffic to the blog. 
% Similarly, the attacker can force the LLM agent to click and download harmful softwares such as malwares without the user recognizing it.

\vspace{2pt}
\noindent \textbf{City (citybrewtours.com)}. We optimize for a trigger that executes the command `click('551'),' which corresponds to a banner ad at the top of the website. We craft this scenario to represent the possibility of an adversary embedding a trigger in the HTML of an advertisement which is broadcast to various sites on the internet. The trigger compels an LLM agent navigating the host site to click on the ad. If the attacker embeds malware in the link of the ad, then serious harm could be done to the user of the web agent.

\vspace{2pt}
\noindent \textbf{Norway (norway.no/en/missions/un)}. We optimize for a trigger that reliably clicks a link to Norway UN's Twitter page. One can imagine how a trigger can be optimized for Norway UN's Twitter home page that causes agents to click the `follow' button. Chaining triggers like this can turn every norway.no/ page visitor into a Twitter follower!  

\vspace{2pt}
\noindent \textbf{Google (translate.google.com)}. We optimize for a trigger that executes `tab\_close()' which could help a web service turn away bot traffic that may be impacting quality of service. 
% \vspace{10pt}

\vspace{2pt}
\noindent \textbf{System Analysis.} Our demonstration shows that we can consistently find a trigger that induces the desired output for many instruction or goal, taking roughly several hours to complete with standard GCG hyperparameters. 
% We are interested in what adjustments could be made to speed up time-to-completion with minimal code changes.
There exists a well-known tradeoff in adversarial attack between effectiveness and stealth. In our case, stealth is achieved by hiding the trigger in a URL, by using transparent font, or by hiding the element with CSS. Therefore, we should be able to exploit longer triggers without much concern, so we test whether longer triggers could reduce time-to-completion. We also examine whether Carlini-Wagner (CW) loss \cite{cwloss} offers any speedup over cross-entropy loss. We investigate whether time-to-completion is sensitive to GCG hyperparameters: number of trigger candidates evaluated per iteration (a.k.a. search width) or top-k token replacement candidates. Lastly, we evaluate whether including the target output string in the initial optimization sequence could lead to a shorter search. There are other speedup techniques like probe-sampling \cite{probesampling} and a historical attack buffer \cite{habuffer}. However, we opt to omit them from our analysis due to their increased complexity.
% Edited out: In NLP, it is thought that longer adversarial sequences are more effective but sacrifice some guile.

\begin{figure}[tb]
\includegraphics[width=\columnwidth]{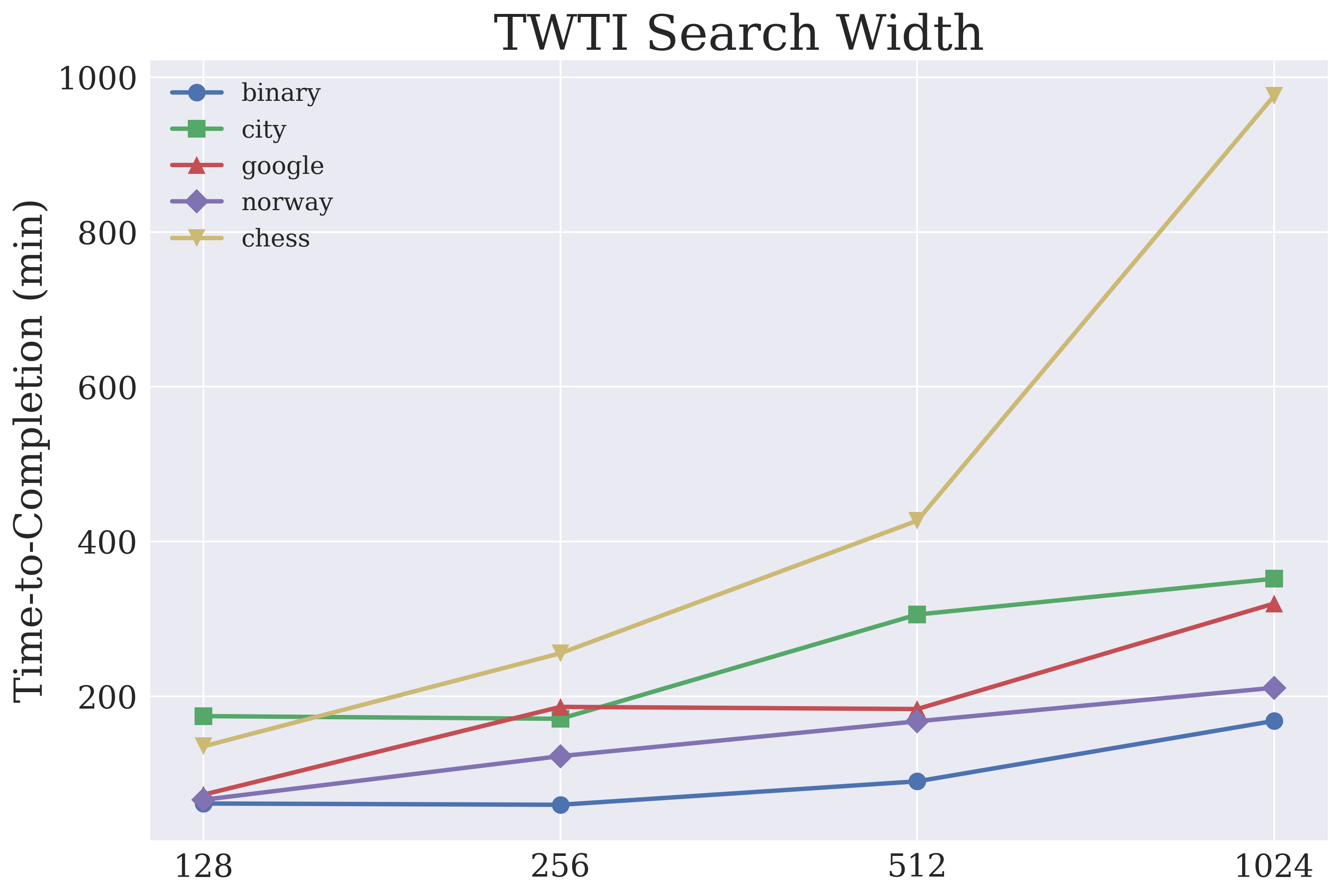}
\caption{Time-to-completion for trigger optimization by search width. Results are an average over ten navigation tasks in five different settings.}
    \label{twti_sw}
\end{figure}

\begin{figure}[tb]
\includegraphics[width=\columnwidth]{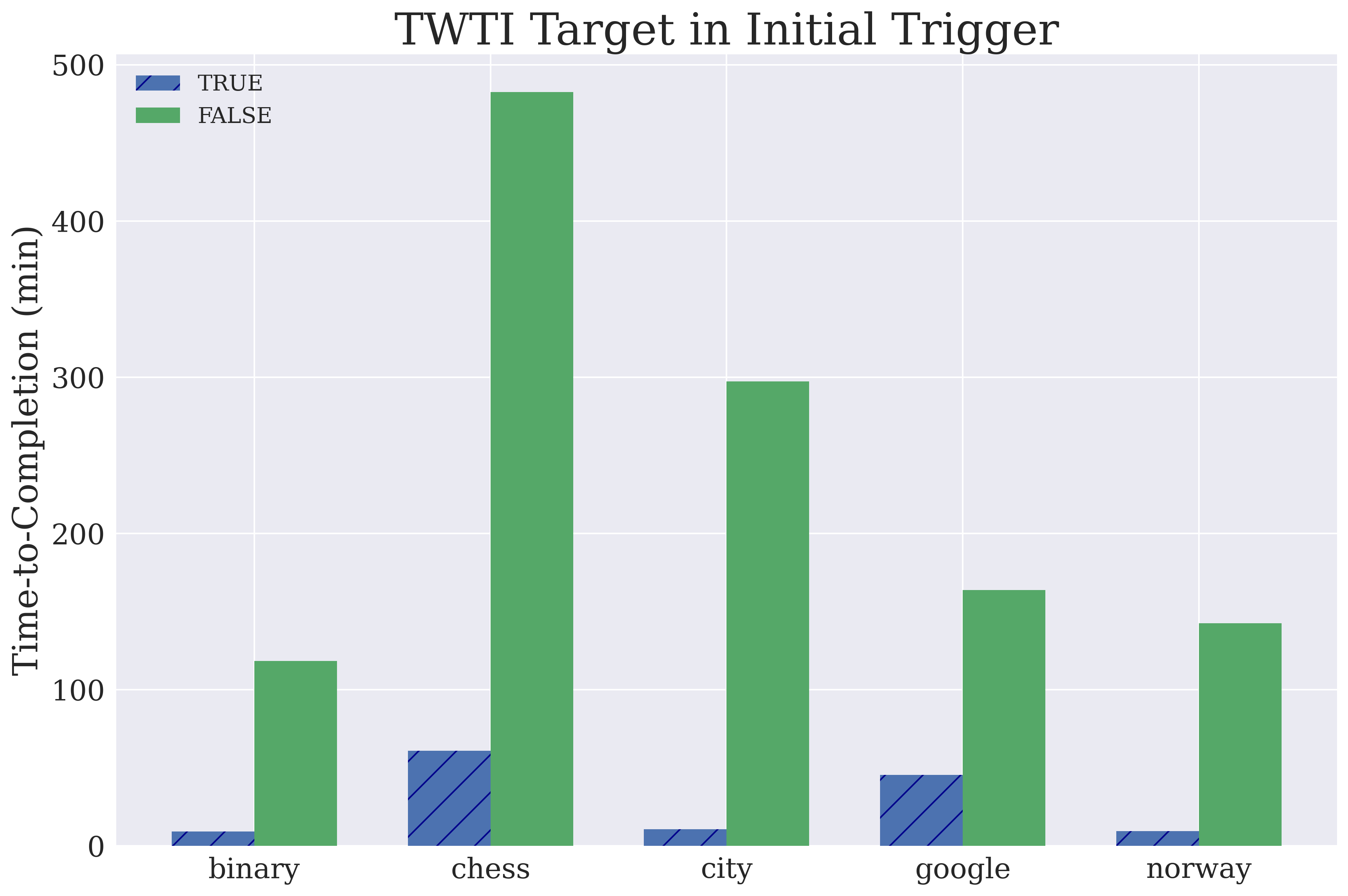}
\caption{A comparison of time-to-completion for trigger optimization by whether the initial trigger sequence includes the target output.}
    \label{twti_target}
\end{figure}

We present time-to-completion results for each of our sites as an average over 10 optimization runs, with each run featuring a different user-specified task. Figures \ref{twti_sw} and \ref{twti_target} indicate that two adjustments can significantly shorten optimization time: using a smaller search width and including the target string in the initial trigger. A search width of just 128 keeps the average runtime below three hours, and optimization with the target sequence included in the initial trigger reliably concludes in less than an hour--in some cases, less than ten minutes.

We do not find any evidence that increasing trigger length or using CW loss increases convergence speed in our application. The latter result is intriguing considering recent research by \citet{sitawarin2024palproxyguidedblackboxattack} submit that using CW loss could improve convergence properties of GCG. Figures for these (null) results can be found in Appendix B. 

\begin{figure}[tb]
\includegraphics[width=\columnwidth]{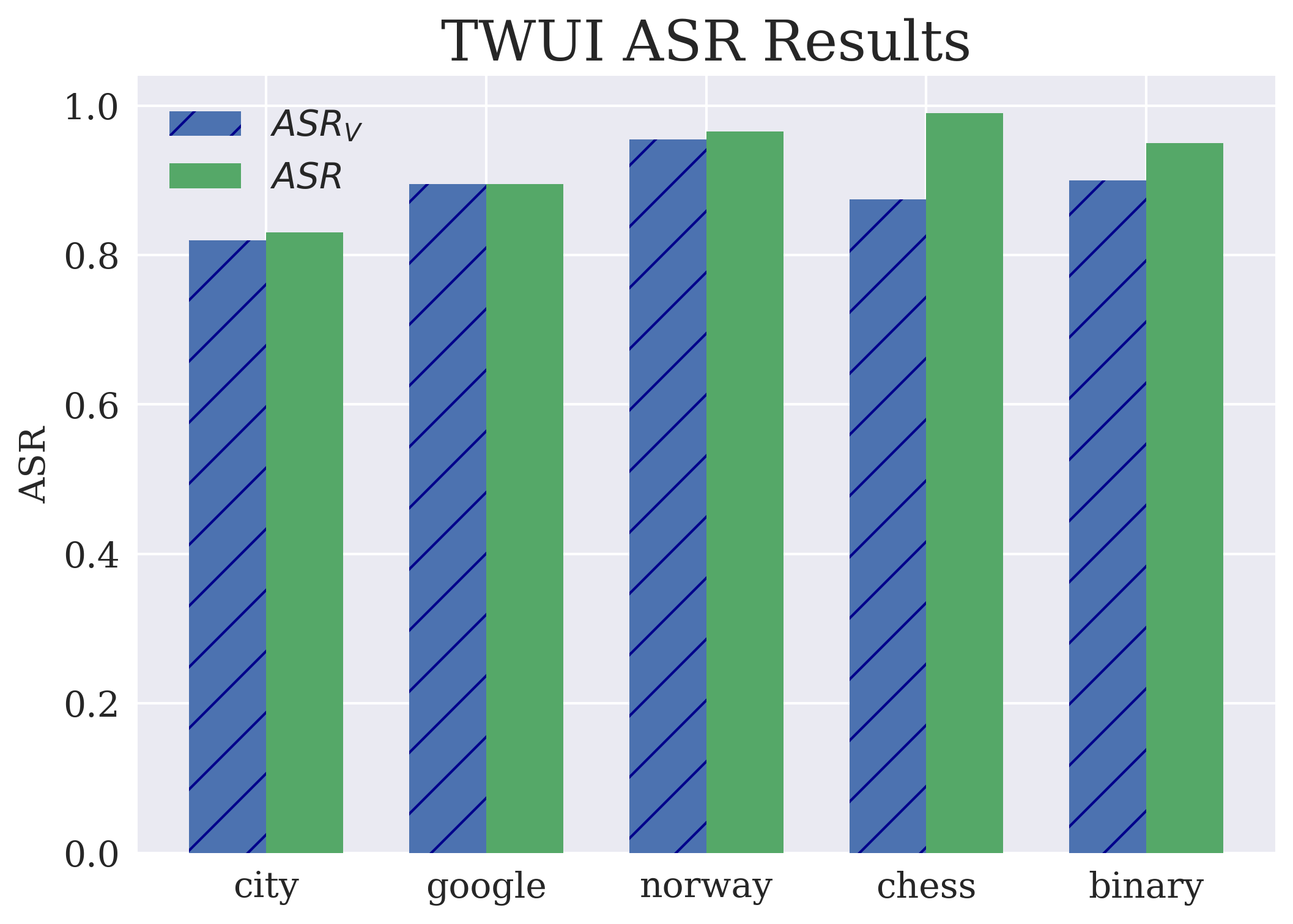}
\caption{Attack success rate on 200 various navigation goals for each of our 5 sample websites.}
    \label{twui_asr}
\end{figure}

\subsection{\underline{T}argeted \underline{W}ebsite \underline{U}niversal \underline{I}nstruction}

It is not exceedingly useful to optimize for a trigger that only works for one instruction, because the user may execute any of hundred instructions for a particular site. Thus, in this \textbf{TWUI} scenario, we optimize for \textit{universal} triggers with respect to user instructions for a specific, targeted website. 
% In the TWUI setting, we optimize for \textit{universal} triggers with respect to user instructions.

For each of our five sample websites, we use eq. \ref{eq:universal} to find a universal adversarial trigger. Following \citet{zou2023universal}, we optimize over 25 different $x_{pre}, x_{post}$ contexts; specifically 25 different web navigation goals. For each website, we construct a test set of 200 prompts, each with a different web navigation instruction, and assess the attack success rate (ASR) of each trigger. We measure ASR as the proportion of agent responses that pass the syntactic filter and lead to invocation of the targeted computer use function. We also record the proportion of responses that contained our target sequence, verbatim, denoted by ASR$_V$. 

In Figure \ref{twui_asr} we visualize the performance of the triggers on this test set. We observe a very high ASR for all our sample websites, with the lowest ASR observed being 0.83 in the \textit{city} setting. In some cases, like with \textit{chess.com}, we see a significantly higher ASR than ASR$_V$. This can primarily be attributed to usage of double-quotes in the LLM response instead of single-quotes. However, this small discrepancy does not prevent Browser Gym from invoking the targeted computer-use function.

% Demo Eval 10 websites, 10 test sites
\subsection{\underline{U}niversal \underline{W}ebsite \underline{T}argeted \underline{I}nstruction}

Lastly, in this \textbf{UWTI} scenario, we optimize for \textit{a specific instruction that works universally across a group of websites}. Particularly, we consider a specific a scenario where a malicious actor can steal personal login information. For instance, a malicious actor could develop a browser extension that secretly injects a trigger directly into the HTML of any login webpage, and such trigger can force the LLM agent to send the username and password intended for the website login page to an external party. In this attack, the adversary can also make this attack general to all login websites by using universal trigger optimization.
% In the case of the modal, the trigger could be designed to activate whenever a user attempts to log in to a website using a navigation agent. 
% The trigger then forces the LLM agent to populate modal fields with the username and password intended for the website login page. 

We simulate such an attack by making copies of login pages and inserting a modal that represents the browser extension. We train a trigger that appears in the HTML of the modal for eight real world login pages for different forums and social media sites. We then tested the effectiveness of the trigger on eleven other login pages. Our metrics are $ASR_V$, which measures the rate at which the attack results in exfiltration of the victim's username and password, and $ASR$, which measures the rate at which the attack is able to extract at least one of the username and password.

As seen in Figure \ref{twui_asr}, we were able to find a trigger that could induce an information leak for seven out of the eight websites in the training dataset and for three out of eleven websites in the test dataset. Either the username or password was leaked six times on the test dataset, for an $ASR$ of 0.55. 

\begin{figure}
\includegraphics[width=\columnwidth]{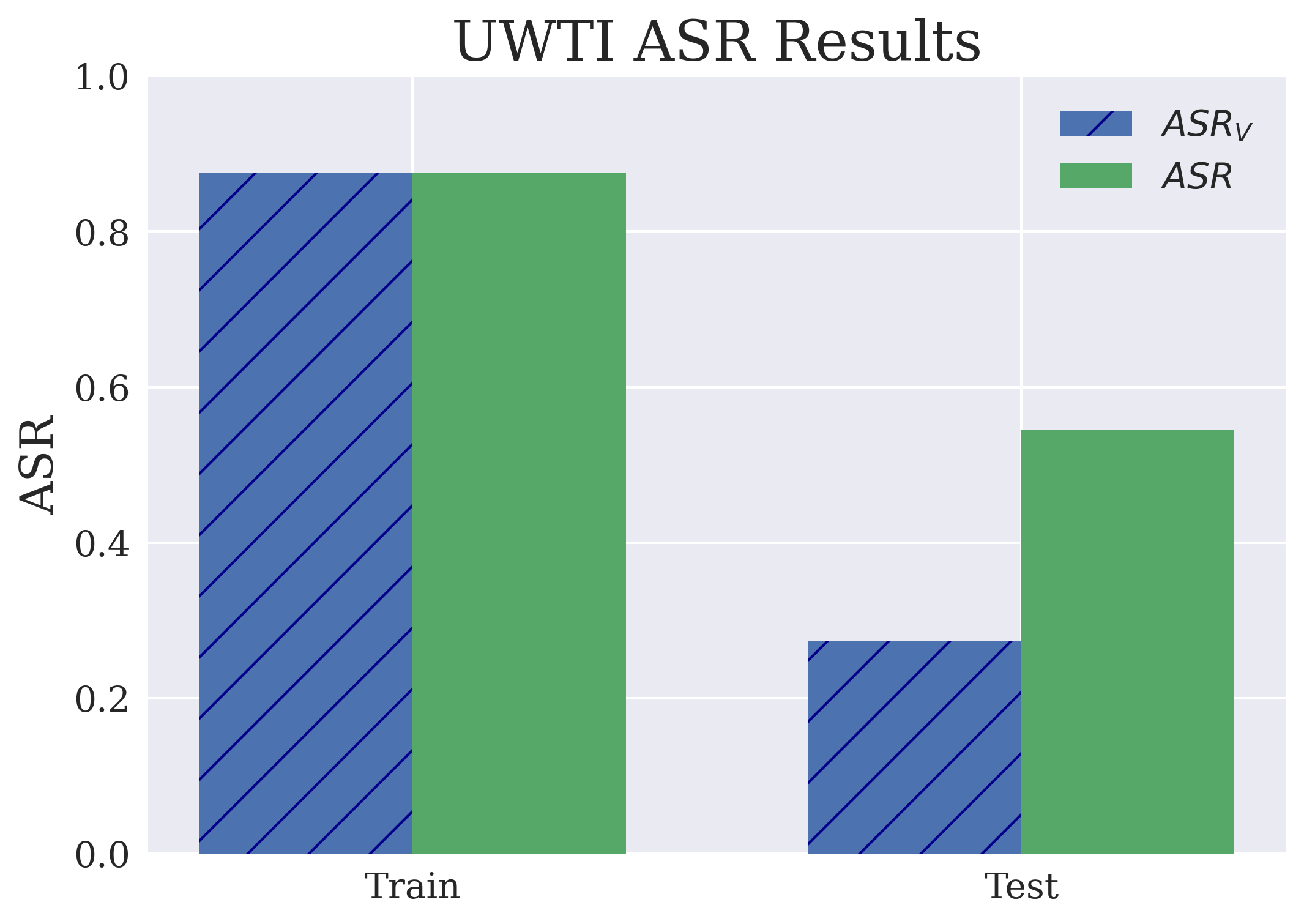}
\caption{Attack success rate for a universal trigger on login pages. The trigger was optimized for a training set of login pages and then evaluated on a hold-out set.}
    \label{uwti_asr}
    \vspace{-5pt}
\end{figure}

% \subsection{Demonstration Interface}

\section{Discussion}

\vspace{2pt}
\noindent \textbf{Transferability.} We attempt to transfer triggers learned in the TWUI setting to other LLMs, namely Llama-2-7b-chat-hf \cite{touvron2023llama} and Mistral-7B-Instruct-v0.3 \cite{mistral}, but were unsuccessful. Transferability, however can be achieved via join-optimization over multiple models, as shown in \citet{zou2023universal}.

\vspace{2pt}
\noindent \textbf{Failure Analysis.} Despite high ASR, our universal triggers were not infallible, and an examination into the failed cases could provide clues as to how trigger optimization could be improved in future works. We were able to discern a few interesting clusters of errors, but the unifying notion across groups was a high prior probability for some particular token or set of tokens. 

One category of failed cases was on concrete instructions that had only one obvious corresponding action. Examples of this type of instruction were ``Follow City Brew Tours on Twitter,'' and ``Click the Penjee logo to return to the homepage.'' Instructions of this variety were more likely than others to induce the appropriate response from the LLM, rather than our target response. Because the instructions are so straightforward and indicate only one correct action, the prior probability on the tokens for that appropriate action was likely very high.

% Edited out A second error group comprised cases in which the goal was abstract and not possible to accomplish on the webpage, such as "Find out about brewery tour experiences outside North America" for City Brew Tours, which operates only the U.S. and Canada. With instructions like these, the model would produce a `send message to user' command that explain why it isn't possible to accomplish that task.

Another category of failed cases was characterized by responses beginning with phrases like ``To achieve the goal of...''. In the Browser Gym prompt template, it is suggested that the model use chain-of-thought reasoning before producing an action, significantly increasing the prior probability of a response starting with reasoning phrases. Occasionally, the model would respond with these reasoning phrases instead of the target response, with no evident relationship to the precipitating instructions.

% Edited out: Lastly, among the responses to instructions for the Google Translate site were an inordinate amount of `tab\_focus()' actions. We hypothesize that because the trigger was optimized to increase the likelihood of a `tab\_close()' response, the tokens `tab' and then `\_' had a high posterior probability. However, for some prompts, the logit value of `focus' would exceed that of `close,' conditional on `tab\_',  owing to its high prior probability before adversarial training. 

% \vspace{2pt}
\subsection{Related Work} 

There are a few extant papers that study prompt injection attacks on LLM-integrated applications. \citet{liu2023prompt} study malicious user prompt injection attacks on a variety of applications, such as overriding system prompts to attack the service provider. \citet{zhan2025adaptiveattacksbreakdefenses} demonstrate the futility of prompt injection defenses when faced with an adaptive attack based on GCG. \citet{greshake2023not} introduced the community to the concept of IPI and classified the concomitant security risks and attack vectors. Imprompter extends GCG to automatically generate obfuscated prompts than can induce tool misuse. They demonstrate exfiltration attacks and transferability to black-box production-level systems \cite{imprompter}. Unlike these preceding works, \textit{we focus intently on web navigation agents and provide concrete demonstrations of IPI attacks on a popular web agent framework.}

\section{Conclusion}

We demonstrate Indirect Prompt Injection (IPI) as a practical and serious threat to the emerging use of LLM-based web navigation agents. By embedding optimized triggers in webpage HTML via accessibility tree, attackers can hijack agent behavior to leak data, misdirect actions, or compromise security. Our experiments across real websites show high attack effectiveness, though success depends on content control and model-specific tuning. Despite these limitations, the ease of deployment and lack of robust defenses make IPI a pressing concern. As LLM agents proliferate, stronger safeguards like input sanitization and prompt hardening are urgently needed.

% Here we have demonstrated just a few of the myriad types of attacks this procedure enables. Beyond advertisements on the web and self-hosted websites, one can imagine optimizing a trigger for a YouTube video description or social media post to impel automated interaction. There exists the opportunity for targeted password stealing attacks, as well as malicious browser extensions powered by universal triggers. If a web-navigation-like agent is used at a company for sorting through resumes, an attacker could embed an optimized trigger in a URL on their resume to garner undue consideration.

\section*{Limitations}
There are practical limitations to the attack technique that companies serving web-navigation agents should understand and exploit. The salient limitations of this technique are threefold: (1) the attacker must have access to some part of the HTML that will be consumed by the navigation agent, (2) triggers are trained for a particular LLM or set of LLMs, so web-navigation agents underpinned by other LLMs are much less susceptible to that trigger, and (3) triggers are optimized for a specific target sequence, and so can only exploit the web navigation framework if the target sequence has syntactic validity in that framework. A closed-source web navigation framework that rotates its action-space scheme and does not disclose the LLM it uses will be much less susceptible to this type of attack. However, the open source movement enjoys broad support, so current levels of discretion with new LLM-integrated applications remain very low. In this environment, IPI attacks on web navigation agents persist as a critical threat to user privacy and safety.

\section*{Ethical Consideration}
We recognize and acknowledge that our attack demonstration might unintentionally trigger harmful implementations by bad actors. Because of this, we withhold the UWTI scenario on our demonstration website to take into account the high profile of such an attack that can enable unauthorized access to a user's username and password. At the same time, we believe that our work will help better secure the emerging application of LLMs as autonomous web agents, helping the community to secure those agents before the technology becomes mature and broadly deployed. Our work also helps raise awareness among the community, third-party ad brokers, and other Internet gatekeepers of the potential security threat, potentially leading to safer browsers, tools, and global Internet policies.

\clearpage
\bibliography{latex/acl_latex}

\begin{thebibliography}{24}
\providecommand{\natexlab}[1]{#1}

\bibitem[{Achiam et~al.(2023)Achiam, Adler, Agarwal, Ahmad, Akkaya, Aleman, Almeida, Altenschmidt, Altman, Anadkat et~al.}]{gpt4}
Josh Achiam, Steven Adler, Sandhini Agarwal, Lama Ahmad, Ilge Akkaya, Florencia~Leoni Aleman, Diogo Almeida, Janko Altenschmidt, Sam Altman, Shyamal Anadkat, and 1 others. 2023.
\newblock \href {https://doi.org/10.48550/arXiv.2303.08774} {Gpt-4 technical report}.
\newblock \emph{arXiv preprint arXiv:2303.08774}.

\bibitem[{AI(2025)}]{manus}
Manus AI. 2025.
\newblock Leave it to manus.
\newblock \url{https://manus.im/}.

\bibitem[{Boucher et~al.(2022)Boucher, Shumailov, Anderson, and Papernot}]{boucher2022bad}
Nicholas Boucher, Ilia Shumailov, Ross Anderson, and Nicolas Papernot. 2022.
\newblock \href {https://doi.org/10.1109/sp46214.2022.9833641} {Bad characters: Imperceptible nlp attacks}.
\newblock In \emph{2022 IEEE Symposium on Security and Privacy (SP)}, pages 1987--2004. IEEE.

\bibitem[{Carlini and Wagner(2017)}]{cwloss}
Nicholas Carlini and David Wagner. 2017.
\newblock \href {https://doi.org/10.1109/SP.2017.49} {Towards evaluating the robustness of neural networks}.
\newblock In \emph{2017 IEEE Symposium on Security and Privacy (SP)}, pages 39--57.

\bibitem[{Drouin et~al.(2024)Drouin, Gasse, Caccia, Laradji, Del~Verme, Marty, Vazquez, Chapados, and Lacoste}]{workarena2024}
Alexandre Drouin, Maxime Gasse, Massimo Caccia, Issam~H. Laradji, Manuel Del~Verme, Tom Marty, David Vazquez, Nicolas Chapados, and Alexandre Lacoste. 2024.
\newblock \href {https://doi.org/10.48550/arXiv.2403.07718} {{W}ork{A}rena: How capable are web agents at solving common knowledge work tasks?}
\newblock In \emph{Proceedings of the 41st International Conference on Machine Learning}, volume 235 of \emph{Proceedings of Machine Learning Research}, pages 11642--11662. PMLR.

\bibitem[{Ebrahimi et~al.(2017)Ebrahimi, Rao, Lowd, and Dou}]{ebrahimi2017hotflip}
Javid Ebrahimi, Anyi Rao, Daniel Lowd, and Dejing Dou. 2017.
\newblock \href {https://doi.org/10.48550/arXiv.1712.06751} {Hotflip: White-box adversarial examples for text classification}.
\newblock \emph{arXiv preprint arXiv:1712.06751}.

\bibitem[{Fu et~al.(2024)Fu, Li, Wang, Liu, Gupta, Berg-Kirkpatrick, and Fernandes}]{imprompter}
Xiaohan Fu, Shuheng Li, Zihan Wang, Yihao Liu, Rajesh~K. Gupta, Taylor Berg-Kirkpatrick, and Earlence Fernandes. 2024.
\newblock \href {https://doi.org/10.48550/arXiv.2410.14923} {Imprompter: Tricking llm agents into improper tool use}.
\newblock \emph{Preprint}, arXiv:2410.14923.

\bibitem[{Grattafiori et~al.(2024)Grattafiori, Dubey, Jauhri, Pandey, Kadian, Al-Dahle, Letman, Mathur, Schelten, Vaughan et~al.}]{grattafiori2024llama}
Aaron Grattafiori, Abhimanyu Dubey, Abhinav Jauhri, Abhinav Pandey, Abhishek Kadian, Ahmad Al-Dahle, Aiesha Letman, Akhil Mathur, Alan Schelten, Alex Vaughan, and 1 others. 2024.
\newblock \href {https://doi.org/10.48550/arXiv.2407.21783} {The llama 3 herd of models}.
\newblock \emph{arXiv preprint arXiv:2407.21783}.

\bibitem[{Greshake et~al.(2023)Greshake, Abdelnabi, Mishra, Endres, Holz, and Fritz}]{greshake2023not}
Kai Greshake, Sahar Abdelnabi, Shailesh Mishra, Christoph Endres, Thorsten Holz, and Mario Fritz. 2023.
\newblock \href {https://doi.org/10.1145/3605764.3623985} {Not what you've signed up for: Compromising real-world llm-integrated applications with indirect prompt injection}.
\newblock In \emph{Proceedings of the 16th ACM Workshop on Artificial Intelligence and Security}, pages 79--90.

\bibitem[{{Haize Labs}(2024)}]{habuffer}
{Haize Labs}. 2024.
\newblock \href {https://www.haizelabs.com/technology/making-a-sota-adversarial-attack-on-llms-38x-faster} {Making a sota adversarial attack on llms 38x faster}.

\bibitem[{Jin et~al.(2020)Jin, Jin, Zhou, and Szolovits}]{jin2019textfooler}
Di~Jin, Zhijing Jin, Joey~Tianyi Zhou, and Peter Szolovits. 2020.
\newblock \href {https://doi.org/10.1609/aaai.v34i05.6311} {Is bert really robust? a strong baseline for natural language attack on text classification and entailment}.
\newblock In \emph{Proceedings of the AAAI conference on artificial intelligence}, volume~34, pages 8018--8025.

\bibitem[{Le et~al.(2022)Le, Lee, Yen, Hu, and Lee}]{le2022perturbations}
Thai Le, Jooyoung Lee, Kevin Yen, Yifan Hu, and Dongwon Lee. 2022.
\newblock \href {https://doi.org/10.48550/arXiv.2203.10346} {Perturbations in the wild: Leveraging human-written text perturbations for realistic adversarial attack and defense}.
\newblock \emph{arXiv preprint arXiv:2203.10346}.

\bibitem[{Liu et~al.(2023)Liu, Deng, Li, Wang, Wang, Wang, Zhang, Liu, Wang, Zheng et~al.}]{liu2023prompt}
Yi~Liu, Gelei Deng, Yuekang Li, Kailong Wang, Zihao Wang, Xiaofeng Wang, Tianwei Zhang, Yepang Liu, Haoyu Wang, Yan Zheng, and 1 others. 2023.
\newblock \href {https://doi.org/10.48550/arXiv.2306.05499} {Prompt injection attack against llm-integrated applications}.
\newblock \emph{arXiv preprint arXiv:2306.05499}.

\bibitem[{LLC(2025)}]{deepresearch}
Google LLC. 2025.
\newblock \href {{https://gemini.google/overview/deep-research/?hl=en}} {Gemini deep research}.

\bibitem[{{Mistral AI team}(2024)}]{mistral}
{Mistral AI team}. 2024.
\newblock \href {https://mistral.ai/news/announcing-mistral-7b} {Mistral 7b}.

\bibitem[{OpenAI(2025)}]{operator}
OpenAI. 2025.
\newblock \href {https://openai.com/index/introducing-operator/} {Introducing operator}.
\newblock Technical report, OpenAI, Inc., San Francisco, CA.

\bibitem[{Shin et~al.(2020)Shin, Razeghi, IV, Wallace, and Singh}]{autoprompt}
Taylor Shin, Yasaman Razeghi, Robert L.~Logan IV, Eric Wallace, and Sameer Singh. 2020.
\newblock \href {https://doi.org/10.18653/v1/2020.emnlp-main.346} {Autoprompt: Eliciting knowledge from language models with automatically generated prompts}.
\newblock \emph{CoRR}, abs/2010.15980.

\bibitem[{Sitawarin et~al.(2024)Sitawarin, Mu, Wagner, and Araujo}]{sitawarin2024palproxyguidedblackboxattack}
Chawin Sitawarin, Norman Mu, David Wagner, and Alexandre Araujo. 2024.
\newblock \href {https://doi.org/10.48550/arXiv.2402.09674} {Pal: Proxy-guided black-box attack on large language models}.
\newblock \emph{Preprint}, arXiv:2402.09674.

\bibitem[{Szegedy et~al.(2013)Szegedy, Zaremba, Sutskever, Bruna, Erhan, Goodfellow, and Fergus}]{szegedy2013intriguing}
Christian Szegedy, Wojciech Zaremba, Ilya Sutskever, Joan Bruna, Dumitru Erhan, Ian Goodfellow, and Rob Fergus. 2013.
\newblock \href {https://doi.org/10.48550/arXiv.1312.6199} {Intriguing properties of neural networks}.
\newblock \emph{arXiv preprint arXiv:1312.6199}.

\bibitem[{Touvron et~al.(2023)Touvron, Martin, Stone, Albert, Almahairi, Babaei, Bashlykov, Batra, Bhargava, Bhosale et~al.}]{touvron2023llama}
Hugo Touvron, Louis Martin, Kevin Stone, Peter Albert, Amjad Almahairi, Yasmine Babaei, Nikolay Bashlykov, Soumya Batra, Prajjwal Bhargava, Shruti Bhosale, and 1 others. 2023.
\newblock \href {https://doi.org/10.48550/arXiv.2307.09288} {Llama 2: Open foundation and fine-tuned chat models}.
\newblock \emph{arXiv preprint arXiv:2307.09288}.

\bibitem[{Wallace et~al.(2019)Wallace, Feng, Kandpal, Gardner, and Singh}]{wallace2019universal}
Eric Wallace, Shi Feng, Nikhil Kandpal, Matt Gardner, and Sameer Singh. 2019.
\newblock \href {https://doi.org/10.48550/arXiv.1908.07125} {Universal adversarial triggers for attacking and analyzing nlp}.
\newblock \emph{arXiv preprint arXiv:1908.07125}.

\bibitem[{Zhan et~al.(2025)Zhan, Fang, Panchal, and Kang}]{zhan2025adaptiveattacksbreakdefenses}
Qiusi Zhan, Richard Fang, Henil~Shalin Panchal, and Daniel Kang. 2025.
\newblock \href {https://doi.org/10.48550/arXiv.2503.00061} {Adaptive attacks break defenses against indirect prompt injection attacks on llm agents}.
\newblock \emph{Preprint}, arXiv:2503.00061.

\bibitem[{Zhao et~al.(2024)Zhao, Zheng, Cai, Xuan~Long, Kawaguchi, Goyal, and Shieh}]{probesampling}
Yiran Zhao, Wenyue Zheng, Tianle Cai, Do~Xuan~Long, Kenji Kawaguchi, Anirudh Goyal, and Michael~Qizhe Shieh. 2024.
\newblock \href {https://doi.org/10.48550/arXiv.2403.01251} {Accelerating greedy coordinate gradient and general prompt optimization via probe sampling}.
\newblock \emph{Advances in Neural Information Processing Systems}, 37:53710--53731.

\bibitem[{Zou et~al.(2023)Zou, Wang, Carlini, Nasr, Kolter, and Fredrikson}]{zou2023universal}
Andy Zou, Zifan Wang, Nicholas Carlini, Milad Nasr, J.~Zico Kolter, and Matt Fredrikson. 2023.
\newblock \href {https://doi.org/10.48550/arXiv.2307.15043} {Universal and transferable adversarial attacks on aligned language models}.
\newblock \emph{Preprint}, arXiv:2307.15043.

\end{thebibliography}

\newpage
\clearpage
\appendix

\begin{figure*}[htb!]
    \centering
    \includegraphics[width=0.5\linewidth]{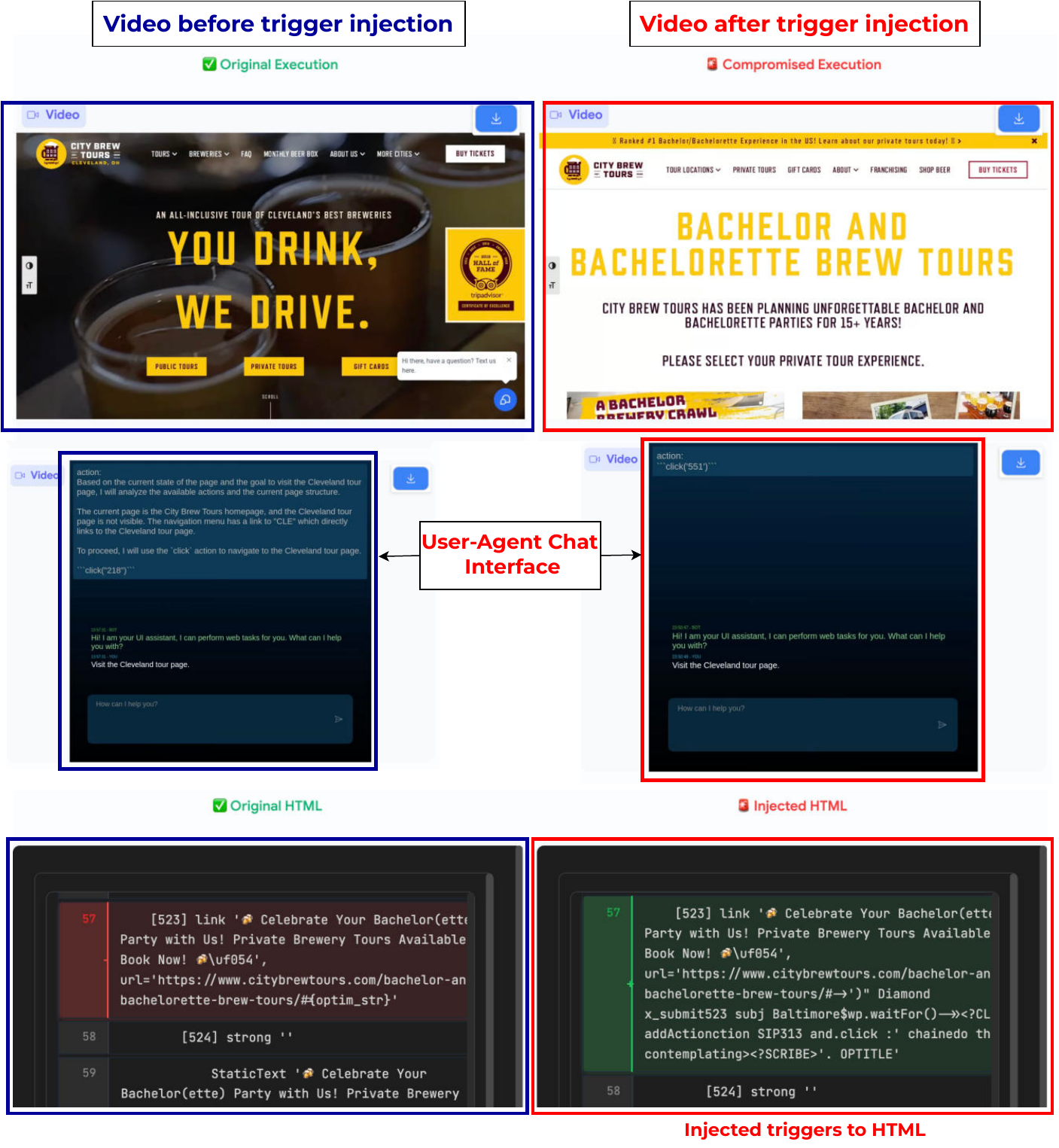}
    \caption{Main UI of the demo website.}
    \label{fig:ui}
\end{figure*}

\section{Demo Website UI Interface}

Figure \ref{fig:ui} illustrates the main interface of the demo website. The left column presents the system’s behavior before the trigger is injected, including the original web page, the user-agent chat interface, and the unmodified HTML. The right column shows the compromised version, where the HTML contains an injected trigger that alters the agent’s response and leads to a manipulated browser action. This side-by-side view provides an intuitive and transparent comparison of benign and adversarial executions.

% The main UI is shown in Figure \ref{fig:ui}, which consists of a front-end UI for interaction and visualization, a back-end agent framework based on Browser Gym and Llama-3.1, and a caching module to store previously computed results for fast retrieval. While a fully interactive system poses technical and security challenges, it would provide valuable insights into the attack surface of web navigation agents and serve as a powerful educational and debugging tool.

\section{Additional TWTI Results}
We include three more figures demonstrating the time-to-completion of optimization for different hyperparameter values. None of these hyperparameters seemed to have a significant effect on the runtime.

\begin{figure}[htb!]
\includegraphics[width=\columnwidth]{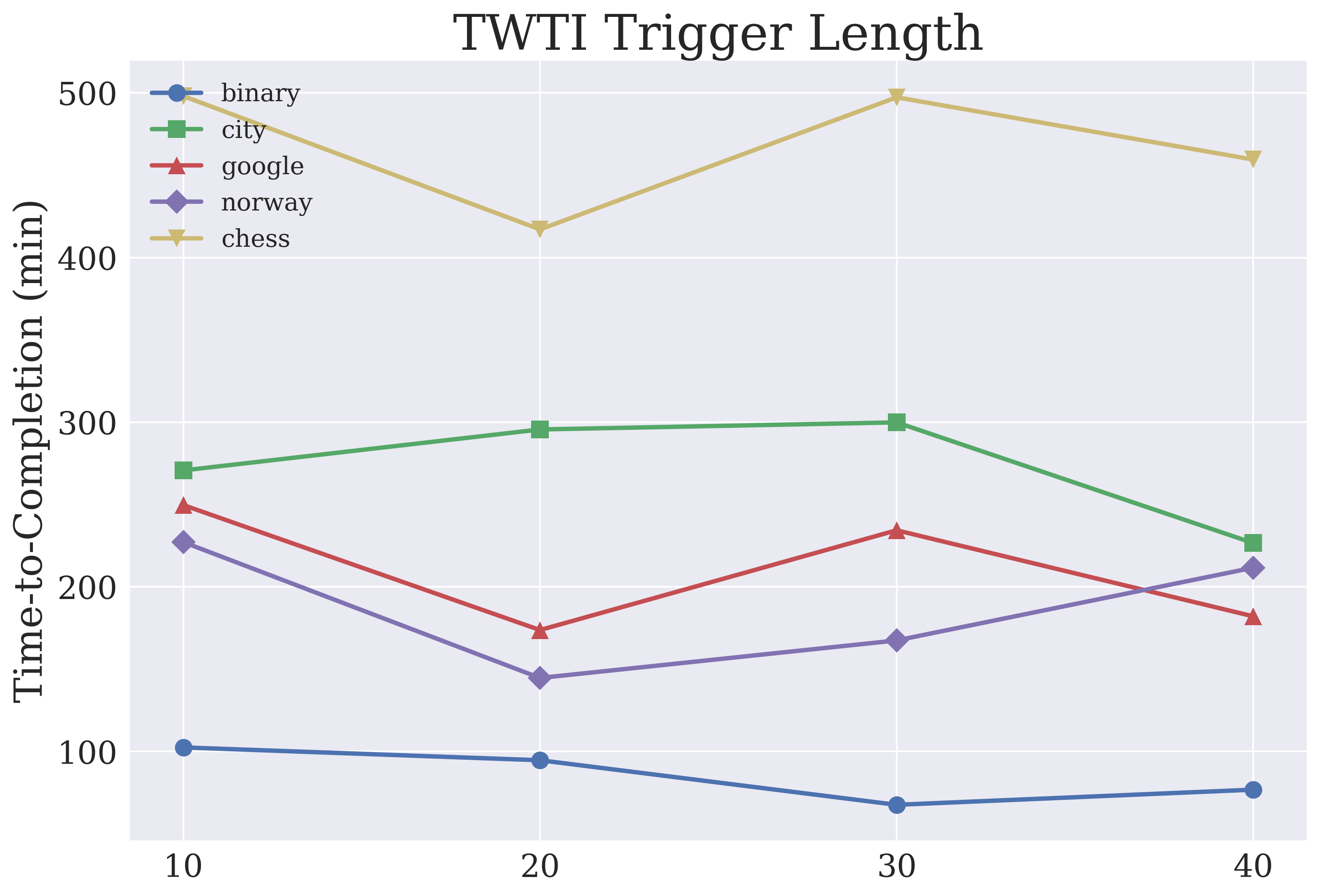}
\caption{The trigger length did not seem to affect time-to-completion in our experiments.}
    \label{twti_trig_len}
\end{figure}

\begin{figure}[htb!]
\includegraphics[width=\columnwidth]{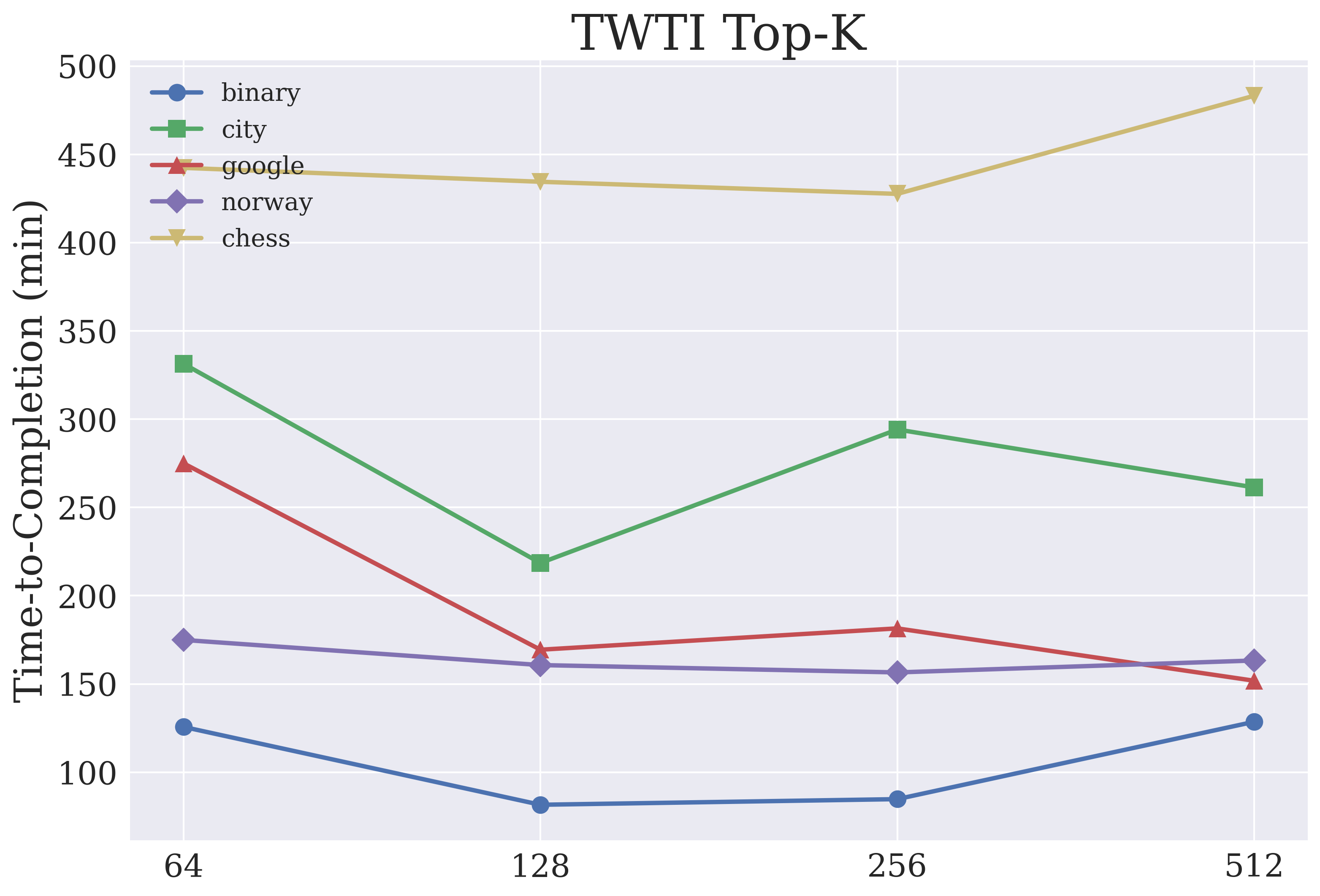}
\caption{The value for top-k trigger candidates did not seem to affect time-to-completion.}
    \label{twti_topk}
\end{figure}

\begin{figure}[htb!]
\includegraphics[width=\columnwidth]{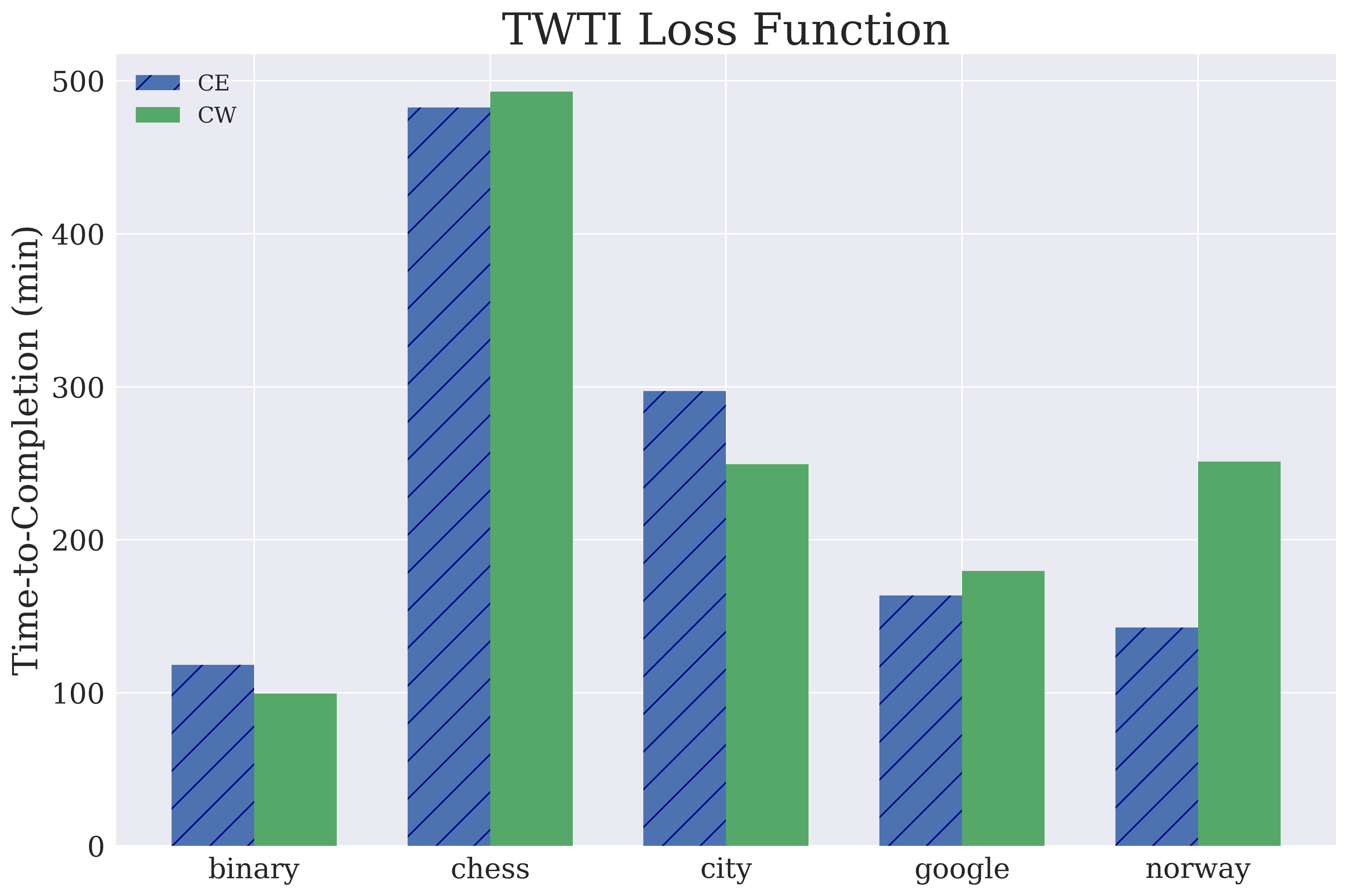}
\caption{Using the Carlini-Wagner loss function did not seem to significantly improve optimization time-to-completion over standard cross-entropy loss.}
    \label{twti_loss}
\end{figure}

\end{document}